\documentclass[epj]{webofc}
\usepackage[varg]{txfonts}   
\wocname{EPJ Web of Conferences}
\woctitle{INPC 2013}
\begin{document}
\title{Fusion using time-dependent density-constrained DFT}
%
%

\author{R. Keser\inst{1,2}\fnsep\thanks{\email{recep.keser@erdogan.edu.tr}} \and
        A.S. Umar\inst{1} \and
        V.E. Oberacker\inst{1} \and
        J.A. Maruhn\inst{3} \and
        P.-G. Reinhard\inst{4}
}

\institute{Physics and Astronomy, Vanderbilt University, Nashville, TN 37235, USA
\and
           RTE University, Science and Arts Faculty, Department of Physics, 53100, Rize, TURKEY
\and
           Institut f\"ur Theoretische Physik, Goethe-Universit\"at, D-60438 Frankfurt am Main, Germany
\and
           Institut f\"ur Theoretische Physik, Universit\"at Erlangen, D-91054 Erlangen, Germany
          }

\abstract{%
 We present results for calculating fusion cross-sections using a new microscopic approach 
 based on a time-dependent density-constrained
 DFT calculations. The theory is implemented by using densities and other information
 obtained from TDDFT time-evolution of the nuclear system as
 a constraint on the density for DFT calculations.
}
\maketitle
\section{Introduction}
\label{intro}
The investigation of internuclear potentials for heavy-ion collisions is
of fundamental importance for the study of fusion reactions
as well as for the formation of
superheavy elements and nuclei far from stability.
Recently, we have developed a new method to extract ion-ion interaction potentials directly from
the time-dependent Hartree-Fock (TDHF) time-evolution of the nuclear system~\cite{UO06b}.
In the density-constrained TDHF (DC-TDHF) approach
the TDHF time-evolution takes place with no restrictions.
At certain times during the evolution the instantaneous density is used to
perform a static Hartree-Fock minimization while holding the neutron and proton densities constrained
to be the corresponding instantaneous TDHF densities. In essence, this provides us with the
TDHF dynamical path in relation to the multi-dimensional static energy surface
of the combined nuclear system. In this approach
there is no need to introduce constraining operators which assume that the collective
motion is confined to the constrained phase space. In short, we have a self-organizing system which selects
its evolutionary path by itself following the microscopic dynamics.
Some of the effects naturally included in the DC-TDHF calculations are: neck formation, mass exchange,
internal excitations, deformation effects to all order, as well as the effect of nuclear alignment
for deformed systems.
The DC-TDHF theory provides a comprehensive approach to calculating fusion barriers
in the mean-field limit. The fusion cross-sections are subsequently calculated by integrating the
Schr\"odinger equation using the incoming-wave boundary condition method (IWBC) as outlined in
Ref.~\cite{UO09a}.
The theory has been applied to calculate fusion cross-sections
for a large number of systems~\cite{UO06c,UO06d,UO07a,UO08a,UO09a,UO10a,OU10}.
In this contribution we will outline the DC-TDHF method and give new examples of its 
application to the calculation of fusion cross-sections for several systems.

\section{Recent Applications of the DC-TDHF Method}
\begin{figure}[!htb]
\begin{centering}
\includegraphics*[height=.263\textheight]{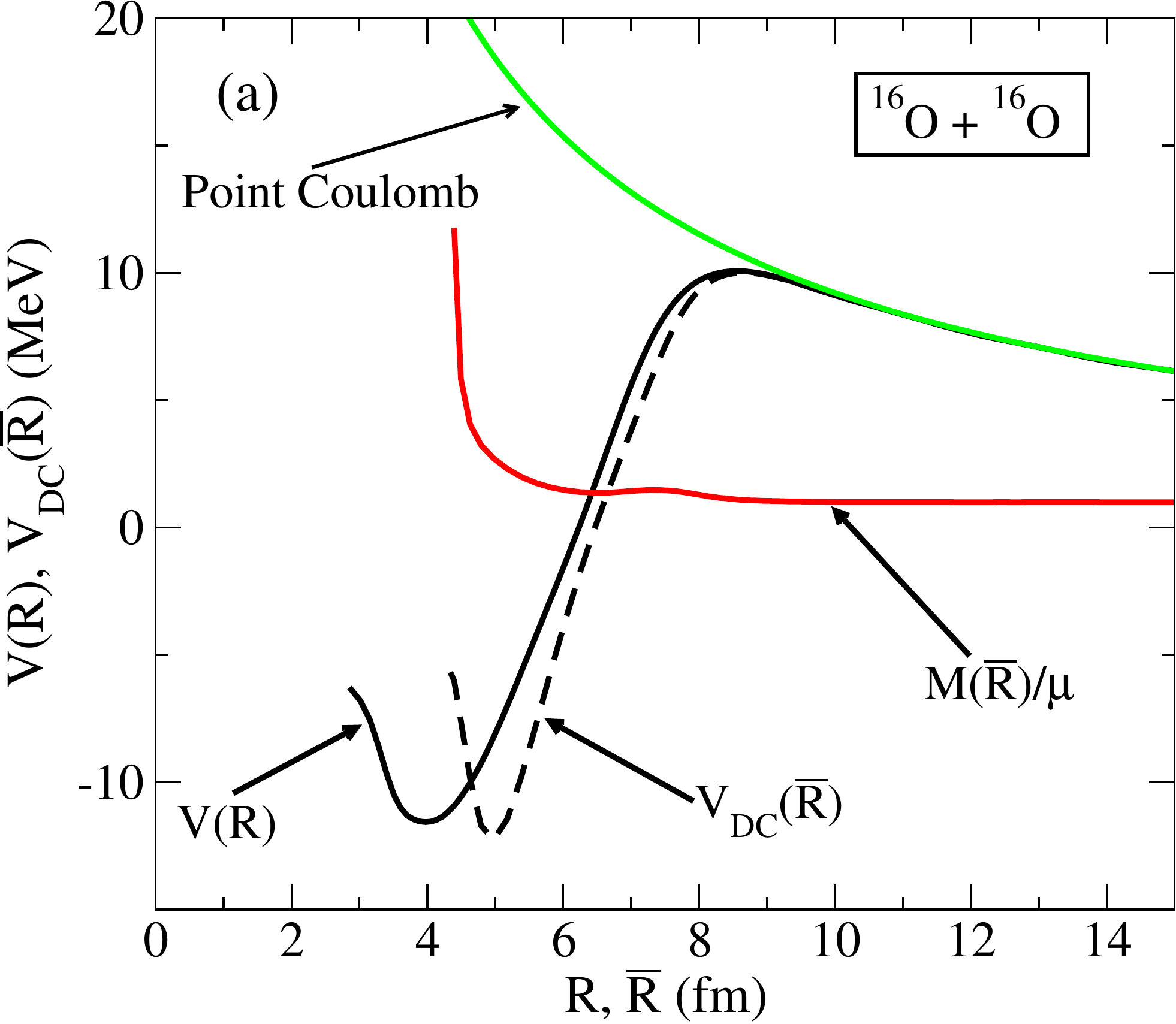}\hspace{0.04in}\includegraphics*[height=.26\textheight]{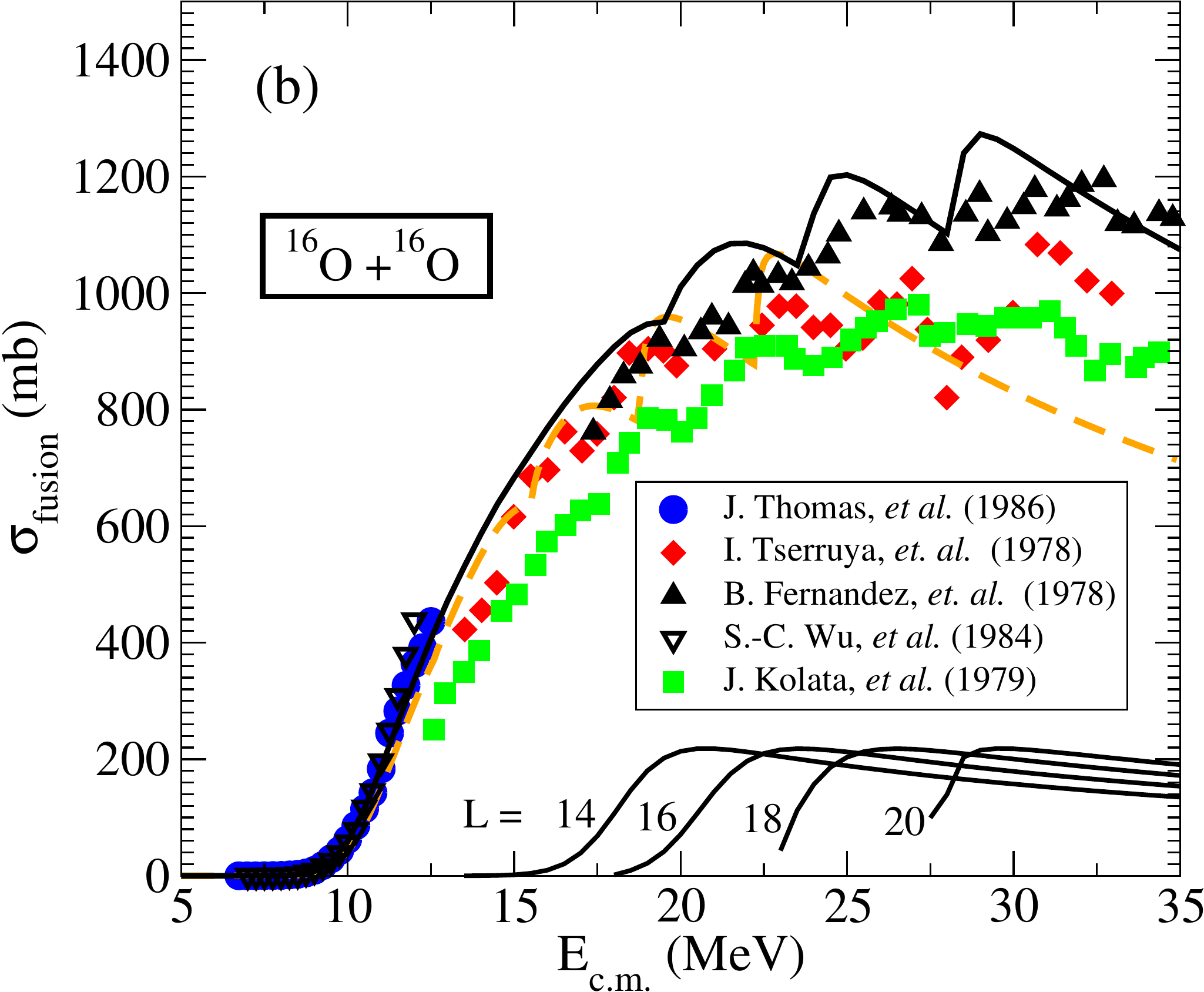}
\caption{\label{fig1} (a) Ion-Ion potential and effective mass for $^{16}$O+$^{16}$O.
                      (b) Corresponding fusion cross-sections.}
\end{centering}
\end{figure}

In this Section we give some recent examples of DC-TDHF calculations of heavy-ion potentials and cross-sections.
Recently, we have studied the fusion of very neutron rich light nuclei that may be important to determine the composition
and heating of the crust of accreting neutron stars~\cite{DC-TDHF-OO}.
The main focus was the O$+$O and C$+$O systems. For the $^{16}$O+$^{16}$O system we have shown
excellent agreement between our calculations and the low energy data from Refs.~\cite{o16o16expdata_1,o16o16expdata_2}.
We have also extended this work to higher energies to see how our results compare with the available
data. The reactions of light systems at high energies ($2-3$ times the barrier height) is complicated
both experimentally and theoretically due to the presence of many breakup channels and excitations.
All the data we could find date back to late 1970's~\cite{Fernandez78,Tserruya78,Kolata79}.
Experimental findings differ considerably in this energy regime as can be seen from Fig.~\ref{fig1}.
Recent analysis of the $^{16}$O+$^{16}$O system system by H.~Esbensen~\cite{EsbensenOO} primarily
uses the data of Tserruya {\it et al.}~\cite{Tserruya78}.
We expect the TDHF results to yield a higher fusion cross-section since many of the
breakup channels are not naturally available in TDHF.
However, a close investigation of the TDHF dynamics and the microscopically calculated
excitation energy clearly indicate that a significant portion of the collective kinetic
energy is not equilibriated. 
It may be plausible to consider the direct influence of the excitation energy, $E^*(R)$,
on the fusion barriers by making an analogy with the coupled-channel approach and
construct a new potential $V^*(R)=V(R)+E^*(R)$, which has all the excitations added
to the ion-ion potential $V(R)$ that should be calculated at higher energies to
minimize the nuclear rearrangements (frozen-density limit).
The resulting potentials somewhat resemble the repulsive-core coupled-channel
potentials of Ref.~\cite{Esb10}.
This approach does
lead to improvements in cases where most of the excitation energy
is in the form of collective excitations rather than irreversible
stochastic dissipation (true especially for lighter systems).
The viability of this approach requires further examination and will be studied in the future.
It is interesting to note that the gross oscillations in the cross-section at higher energies
are correctly reproduced in our calculations. This is simply due to opening of new $L$-channels
as we increase the collision energy. Individual contributions to the cross-section
from higher $L$ vales are also shown on the lower part of the  plot.
\begin{figure}[!htb]
\begin{centering}
\includegraphics*[height=.263\textheight]{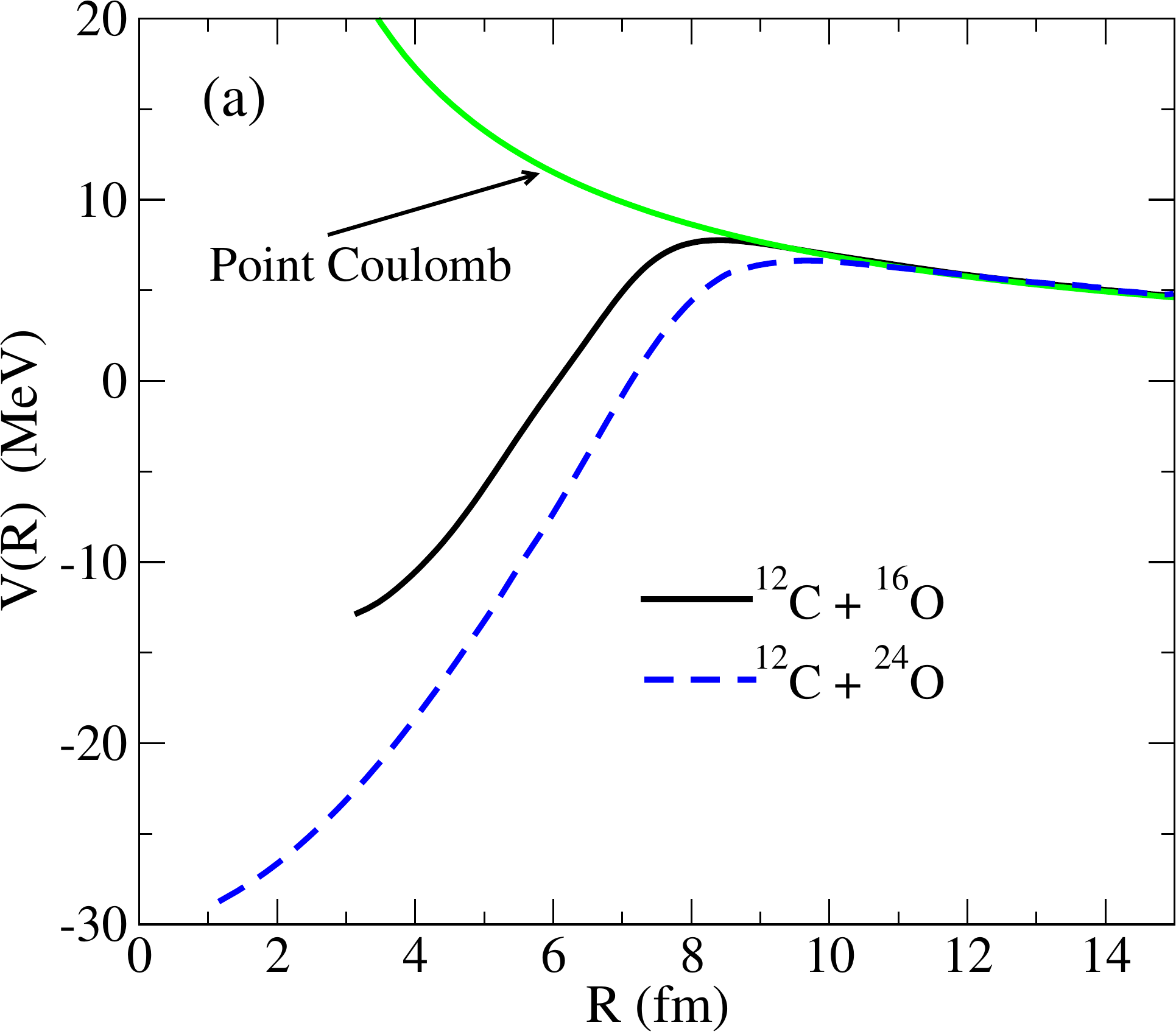}\hspace{0.04in}\includegraphics*[height=.26\textheight]{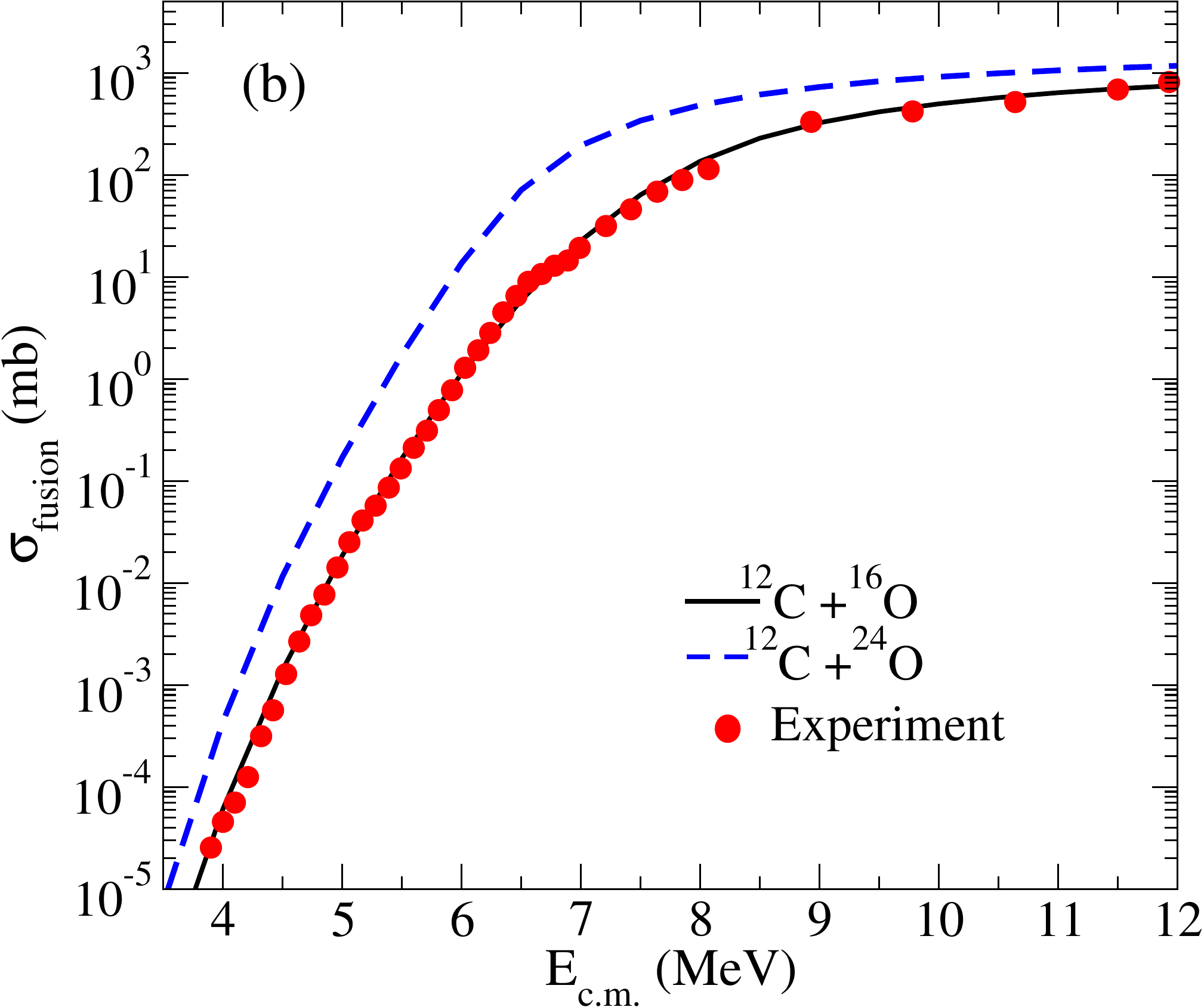}
\caption{\label{fig2} (a) Ion-Ion potential for various isotopes of the C$+$O system.
                      (b) Corresponding cross-sections.}
\end{centering}
\end{figure}

In Fig.~\ref{fig2}a we show the DC-TDHF potential barriers for the C$+$O system.
The higher barrier corresponds to the $^{12}$C$+$ $^{16}$O system and has a peak
energy of $7.77$~MeV. The barrier for the $^{12}$C$+$ $^{24}$O system occurs at a
slightly larger $R$ value with a barrier peak of $6.64$~MeV.
Figure~\ref{fig2}b shows the corresponding cross sections for the two reactions.
Also shown are the experimental data from Refs.~\cite{c12o16expdata_1,c12o16expdata_2,c12o16exp}. The
DC-TDHF potential reproduces the experimental cross-sections quite well for the
$^{12}$C$+$ $^{16}$O system, and the cross section for the neutron rich $^{12}$C+$^{24}$O
is predicted to be larger than that for $^{12}$C+$^{16}$O.
\begin{figure}[!htb]
\begin{centering}
\includegraphics*[height=.26\textheight]{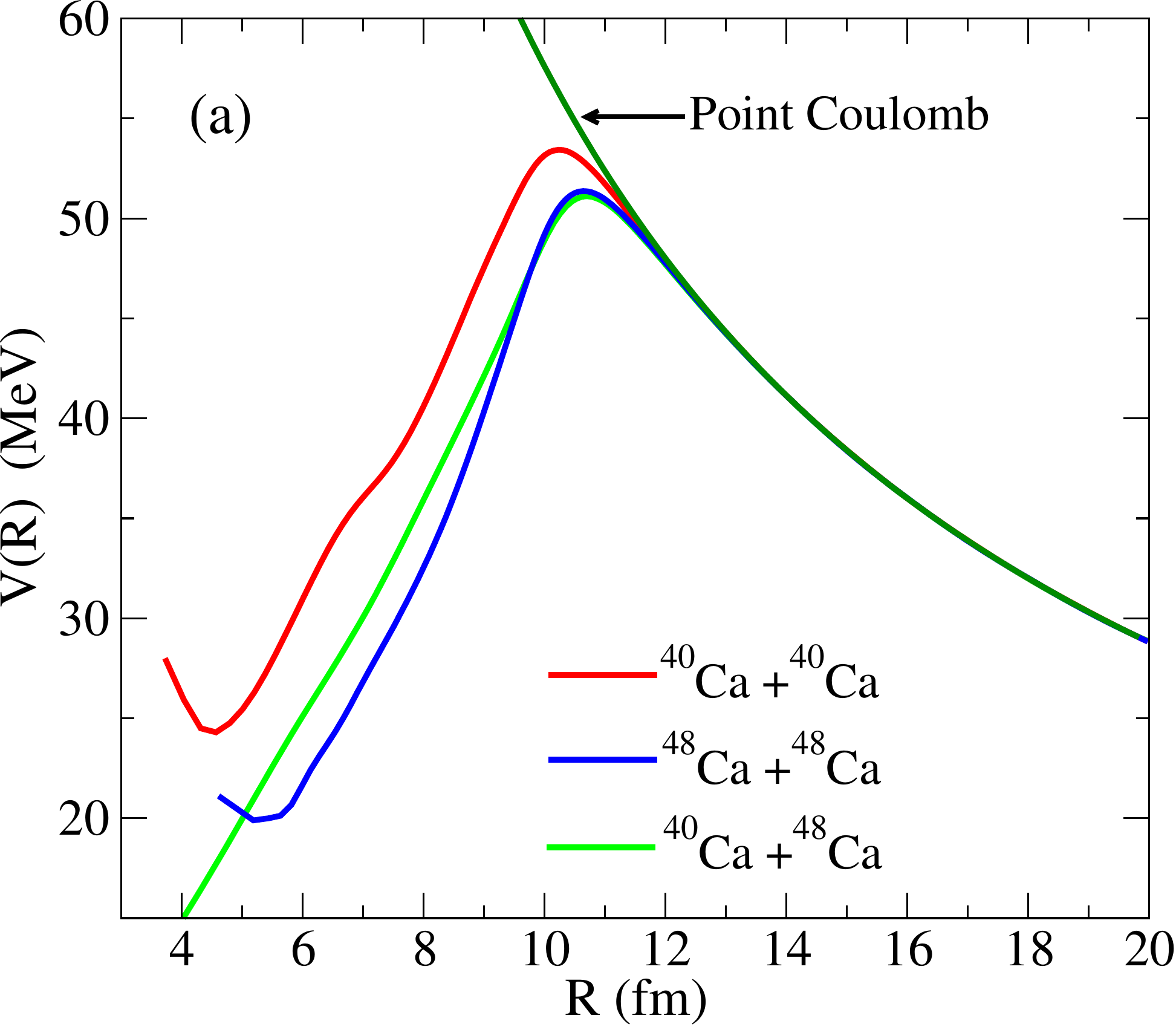}\hspace{0.04in}\includegraphics*[height=.265\textheight]{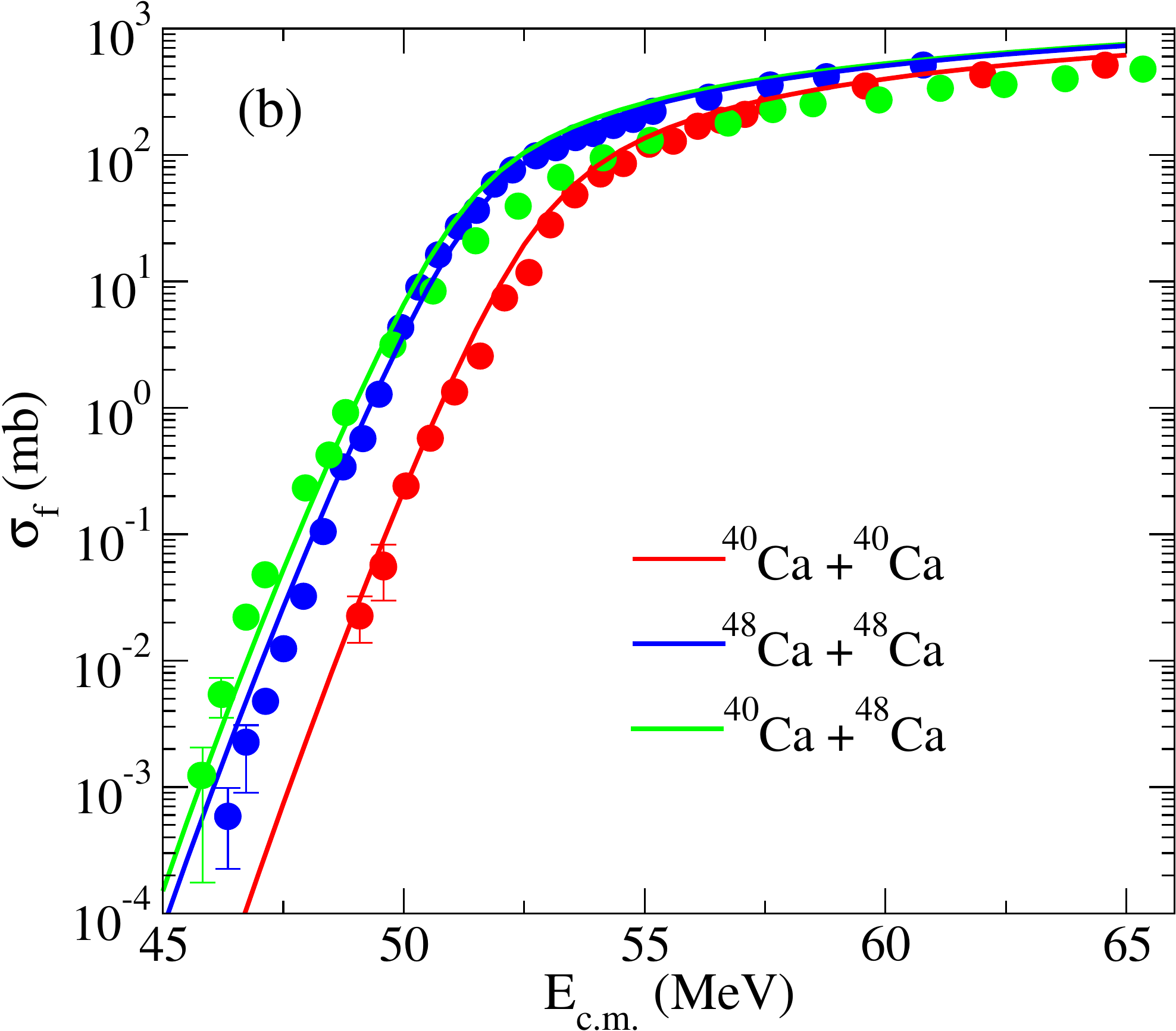}
\caption{\label{fig3} (a) Ion-Ion potential for various isotopes of the Ca$+$Ca system.
                      (b) Corresponding cross-sections.}
\end{centering}
\end{figure}

Figures~\ref{fig3}a and \ref{fig3}b show the corresponding potentials and cross-sections for the Ca$+$Ca system~\cite{DC-TDHF-Ca},
which was the subject of recent experimental studies~\cite{Mon11}.
The observed trend for sub-barrier
energies is typical for DC-TDHF calculations when the underlying microscopic interaction
gives a good representation of the participating nuclei. Namely, the potential barrier
corresponding to the lowest collision energy gives the best fit to the sub-barrier cross-sections
since this is the one that allows for more rearrangements to take place and grows the inner part
of the barrier. Considering the fact that historically the low-energy sub-barrier cross-sections
of the $^{40}$Ca+$^{48}$Ca system have been the ones not reproduced well by the standard models,
the DC-TDHF results are quite satisfactory, indicating that the dynamical evolution of the
nuclear density in TDHF gives a good overall description of the collision process.
The shift of the cross-section curve with increasing collision energy is typical.
In principle one could perform a DC-TDHF calculation at each energy above the barrier
and use that cross-section for that energy. However, this would make the computations
extremely time consuming and may not provide much more insight.
The trend at higher energies for the $^{40}$Ca+$^{48}$Ca system is atypical. The calculated cross-sections
are larger than the experimental ones by about a factor of two.
Such lowering of fusion cross-sections with increasing collision energy
is commonly seen in lighter systems where various inelastic channels,
clustering, and molecular formations are believed to be the contributing
factors.

\section*{Acknowledgments}
This work has been supported by the U.S. Department of Energy under grant No.
DE-FG02-96ER40975 with Vanderbilt University. RK thankfully acknowledges The Scientific and
Technological Council of Turkey (TUBITAK, BIDEB-2219) for a partial support.

\end{document}